\documentclass[twocolumn,twoside,slac_two,nofootinbib]{revtex4-1}
\usepackage{graphicx}
\usepackage{fancyhdr}
\usepackage{hyperref}
\usepackage{graphicx}
\usepackage{amssymb}
\usepackage{verbatim}
\usepackage{multirow}
\usepackage{tabularx, colortbl}
\usepackage{amsmath,esint}

\hypersetup{
    colorlinks=true,
    urlcolor=magenta,
}
\def\svu{cm$^{3}$~s$^{-1}$}
\def\sv{$\langle \sigma v \rangle$}

\let\oldenumerate\enumerate
\renewcommand{\enumerate}{
  \oldenumerate
  \setlength{\itemsep}{1pt}
  \setlength{\parskip}{0pt}
  \setlength{\parsep}{0pt}
}
\let\olditemize\itemize
\renewcommand{\itemize}{
  \olditemize
  \setlength{\itemsep}{1pt}
  \setlength{\parskip}{0pt}
  \setlength{\parsep}{0pt}
}

\begin{document}

\title{A review of the past and present MAGIC dark matter search
  program \\ and a glimpse at the future}

%

\author{Michele Doro\\
\small{on behalf of the MAGIC Collaboration}}
\affiliation{University \& INFN Padova, via Belzoni 7, 35131 Padova,
  Italy, michele.doro@pd.infn.it}

\begin{abstract}
The MAGIC TeV gamma-ray telescopes have devoted several hundreds hour
of observation time in about a decade, to hunt for particle dark matter 
indirect signatures in gamma rays, from various candidate targets of interest
in the sky: the galactic center, satellite galaxies, galaxy clusters and
unidentified objects in other bands. Despite the effort, no hints are
present in MAGIC data. These observation are nevertheless not
unusable. MAGIC indeed derived the most robust upper limits in the TeV
range than any other instrument. These results, for the time being, only mildly
constrain some classic dark matter models, but are of use in the
construction of dark matter models for the next searches, that
consider also the negative results from accelerator and
direct-detection experiments.

In the contribution, we discuss and review MAGIC results, putting them
into context, and in perspective with the next generation of
ground-based Cherenkov telescopes. We will briefly inform about future
MAGIC projects regarding dark matter searches.
\end{abstract}

\maketitle

\thispagestyle{fancy}

\section{\label{sec:intro}Introduction}
The understanding of the nature of Dark Matter (DM), either as a new
particle~\cite{Feng:2010gw} or as a modification of the gravitational law~\cite{Milgrom:1983ca}, is keeping hundreds
of scientists and instruments occupied in hunting for
significant signatures, especially in the past two decades. The need of
DM appears from several observations, all connected to gravitational
effects, at all cosmological scales: from the galactic motion of stars,
to that of galaxies in clusters, and farther away into the signatures
in acoustic oscillations in the cosmic microwave background~\cite{Ade:2015xua}. It is
very hard to disprove such strong hints of gravitational imbalances,
as well as it is easier to explain those with the introduction of one
(or more) new particle(s), to be the candidate for DM. This new
particle should be either stable, or very long lived, to guarantee the
relic density we see today $\Omega_{DM}=0.259\pm 0.006$~\cite{Ade:2015xua}. This one
parameter, allows anyhow for a very large parameter space in terms of
DM particle mass and annihilation or decay rates. However, the current
paradigm focuses on a ``cold'' DM scenario, in which the particle has
been non relativistic since its decoupling. The velocity of DM particles
determines their free-streaming length, and thus strongly affect the
cosmic structures formation, at least in the current preferred
bottom-up scenario of merging of smaller structures into larger
ones. The best CDM candidate is a particle that does not have any
standard interaction, and is a WIMP (Weakly Interacting
Massive Particles). The WIMP may live in a dark sector (no interaction
with the standard model particles), which would make the detection
prospects difficult, or have some channels to the standard models, like in
the case of Super-symmetrical extensions of the standard models (SUSY),
or Unified Extra Dimensions theories. The non-observation of such
particle at LHC, albeit with maximum luminosity, has pushed the searches
to a mass range of several tens or hundreds of GeV for the mass of the
particle~\cite{Feng:2013pwa}.

In this energy range, ground based Imaging Atmospheric Cherenkov Telescope
Arrays (IACTAs), observing the Cherenkov light produced in Extended
Atmospheric Showers (EAS) generated in the stratosphere by cosmic
gamma rays, are an optimal instrument to search for WIMPs, for the
following reasons: $a)$ in many scenarios, WIMPs produce a strong
gamma-ray yield in the annihilation or decay products, $b)$ the
neutrality of photons guarantee that telescopes can be pointed to
astronomical places where DM is expected, $c)$ annihilation or decay
spectrum would be universal, and therefore multiple observations of
such spectra from multiple sources would provide a strong claim, and finally $d)$ the
annihilation or decay spectra often show like a cutoff connected to the DM mass, and sometimes
peculiar bumps that could disentangle its origin from a standard
astrophysical one.

The Major Atmospheric Gamma-ray Imaging Cherenkov (MAGIC) instrument
of the IACTA class, is a pair of 17-m diameter parabolic dishes,
operating since 2009 at the Observatorio Roque de Los Muchachos (ORM,
La Palma, Canary Islands, Spain) at 2,200 m. a.s.l, in the Northern
Hemisphere\footnote{MAGIC started its operations in 2004 as
a single telescope.}.
MAGIC operates between few tens of GeV and several tens of
TeV, with a peak sensitivity of $0.66\%$ of the Crab Nebula flux above
280~GeV, an energy resolution of $10-20\%$ and an angular resolution
below $0.1^\circ$~\cite{Aleksic:2014lkm}. MAGIC observes during night time, with
about 1,500~h of available time per year (1,000~h with moonless
night). MAGIC, as well as other IACTA in the field, is a wide-scope
detector, focused mainly on galactic and extragalactic astrophysics
with gamma rays, but with good potential also as particle detectors
(electron/positron, proton/antiproton, and perhaps neutrino) and
exotic physics detectors (see~\cite{Doro:2016scineghe} for a recent
review). MAGIC has devoted a significant fraction of its observation
time, reaching now several hundreds of hours, to DM-related searches at several
candidate targets. These include the Milky Way (MW) barycenter region, dwarf
satellite galaxies (DSGs) orbiting in the MW DM halo, unidentified
Fermi-LAT objects, galaxy clusters, and others (see also
\cite{Doro:2014pga} for a recent collection of DM searches from IACTA
in the past decade). In these 10 years, the preferred targets,
campaigns duration
and data reconstruction have changed. In this contribution, we put the
MAGIC search into context, showing the evolution of this science
field, and concluding with some remarks on the possible future
extension of the MAGIC DM program and the validity of its contribution
in the field.

The gamma-ray flux from DM annihilation or decay at Earth 
can be factorized as a product of a particle-physics factor depending
on the nature of the DM, and an astrophysical factor, depending on
the target distance and DM distribution, and reads as: 
\begin{equation}\label{eq:dm_flux}
\Phi^\gamma=\frac{\widehat{P}\;N^\gamma}{4\pi k\; m^k}\cdot \iint_{\Omega,los}\rho^k\;ds\,d\omega 
\end{equation}
\begin{eqnarray}
\mbox{Annihilating DM} &:& \widehat{P}=\langle \sigma v \rangle;\quad
k=2; \nonumber \\
\mbox{Decaying DM} &:& \widehat{P}=\tau^{-1};\quad k=1; \nonumber
\end{eqnarray}
where $\langle\sigma v\rangle$ is the velocity average DM annihilation
rate and $\tau$ is DM particle lifetime, $N^\gamma$ is the number of
photons promptly produced during an annihilation or decay event, and
the integrals in the astrophysical factor run over the angular
extension of the searched region $\Omega$, and along the line of sight. The
term $\rho$ represents the DM density. 

\section{\label{sec:early}Early searches}
Before the launch of the Fermi-LAT gamma-ray satellite-borne
instrument (at the time called GLAST), back in 2004, there were great
expectations about both ground-based and satellite-borne gamma-ray
instruments to detect gamma rays from DM annihilation in a reasonable
observation time, at least in some optimistic scenarios~(see, e.g.,
\cite{Bergstrom:2005qk}). Table~\ref{tab:targets} reports the full
list of MAGIC DM targets from the time of this early estimation until
present times. Observations are ordered by telescopes setup, and year
and the table provides information about the source class, the observation time, whether
results were discussed in terms of annihilating or decaying DM scenarios, as well
as links to references. However, it is not straightforward to draw
conclusions based only on the observation time devoted to specific
targets in the table. This is due to the fact that MAGIC performance
evolved significantly with time. MAGIC started its operations in 2004 as
single telescope, and was coupled with a second telescope only in
2009. Both the first and second MAGIC telescopes had undergone major upgrades
along the years, that have substantially changed their performance: from
the first single-telescope setup, to the current, MAGIC has improved its
sensitivity by a factor of 4 (meaning a factor of 16 less time
required) at 300~GeV, and a factor of 10 at 50~GeV. In the following,
we discuss the early searches by target class.

\begin{table*}[h!t]
\centering
\begin{tabular}{l|l|l|c|c|cc|c|l}
\hline
{\bf MAGIC} & {\bf Class} & {\bf Target} & {\bf Year} & {\bf Obs. Time} & {\bf Ann.} &
{\bf Decay} & {\bf Ref.} & {\bf Comments}\\
\hline
\hline
{\bf Mono}
& MW   & Galactic Center & 2006/07 & 25   & - & - & \cite{Albert:2005kh}&\\
& DSG  & Draco           & 2007    & 7.8  & X & - & \cite{Albert:2007xg}&\\
&      & Willman~1       & 2008    & 15.5 & X & - & \cite{Aliu:2008ny}&\\
&      & Segue~1         & 2008/09 & 29.4 & X & - & \cite{Aleksic:2011jx}&\\
& Unid & 3EG1835         & 2007    & 25   & X & - & \cite{Doro:2007kta}&\\
& GC   & Perseus         & 2008    & 24.4 & - & - & \cite{Aleksic:2009ir}&\\
& CR   & All-electrons   & 2009/10 & 14   & - & - & \cite{BorlaTridon:2011dk}&\\
\hline\hline
{\bf Stereo}
& Unid & Many            & 2009/12 & 71.3 & F & - & \cite{Satalecka:2015yxa} & Paper in prep.\\
& DSG  & Segue~1         & 2010/13 & 158  & X & X & \cite{Aleksic:2013xea}&\\
& GC   & Perseus         & 2009/14 & 253  & F & F & \cite{Palacio:2015nza}& Paper in prep.\\
& CR   & All-electrons   & 2012/14 & 40   & - & - & \cite{Mallot:tevpa}&\\
&      & Positrons       &         &      & F & - & \cite{Colin:2011wc}&Paper in prep.\\
& MW   & Galactic Center & 2012/16 & 67  & F & F & \cite{Ahnen:2016crz}& Paper in prep.\\
\hline\hline
\end{tabular}
\caption{\label{tab:targets} Compound of observational targets for indirect dark
matter searches with MAGIC. Observations are grouped in two blocks for
MAGIC results when in single-telescope (mono) configuration and as telescope-pair
(stereo). Classes are: MW (Milky Way), DSG (Dwarf Satellite Galaxy),
Unid (Unidentified HE source), GC (Galaxy Cluster), CR (Cosmic
Ray). The observation time is given in hours. For the DM models, ``X''
means the reference provides constraints on that model, ``-'' that the
reference does not discuss that model, ``F'' means that constraints
about that model are foreseen.}
\end{table*}

\bigskip
Dwarf Satellite Galaxies (DSGs) are small galaxies with a common mass
scale~\cite{Strigari:2008ib} commonly believed to be originated in
 DM overdensities present in the MW DM halo. DSGs have a small
stellar content, and especially are almost depleted of gas, showing no or
little stellar activity in the past Gy. Their star velocity
distribution normally hints to large DM content, with mass-to-light
ratio of $1000~$M$_\odot/$L$_\odot$ or even more, depending on the
target object. They are optimal DM
targets because of no expected astrophysical radiation and short
distance. As such, DSGs have been dominating the MAGIC DM observation
program since the beginning. One can easily see that the first MAGIC DM observations, with the
single-telescope, were devoted mostly to short (less than 30h)
observation of DSGs: Draco, Willman~1 and Segue~1. Draco was
considered one of the best candidate at that time, because of its large
mass-to-light ratio, and precise estimation of the stellar motion
thanks to a sample of thousands of star members. It was observed for
about 8h. In 2007, the Sloan Digital Sky Survey started to produce
high-precision photometric data on a newly discovered class of DSG,
the so-called ``ultra-faint'', due to the fact that the number of
member stars were a factor of 10 less than the previously discovered
``classical'' DSG (including Draco). Two ultra-faint targets were
observed: Willman~1 and Segue~1.
The latter is an ultra-faint DSG, largely debated in the literature. It currently comprises about 70
member stars. The result of the J-factor estimation based on Jeans
analysis goes from ranking Segue~1 as the best candidate~\cite{Essig:2010em} to an
unreliable candidate~\cite{Bonnivard:2015xpq}. MAGIC is
recently computing the J-factors using the public CLUMPY
code~\cite{Bonnivard:2015pia} and the conclusions is that Segue~1 is
still one of the candidate with the largest J-factor.
MAGIC devoted again a
limited observation time to these two targets: about 16h for Willman~1 and about 30h for the Segue~1. As a result, we put upper limits
at a level of \sv=$10^{-22}$\svu. The collection of some of the
discussed MAGIC upper limits is shown in Fig.~\ref{fig:evolution}.

Luminous structures like star or gas clouds are believed to form by
gravitational contraction onto primordial DM overdensities. In this
sense, ``light traces matter'', however, several cases were discussed
where DM overdensities could be almost completely dark at all
wavelengths but in gamma-rays, due to negligible stellar activity. This
was the case that motivated our search for Intermediate Mass Black
Holes (IMBHs), following~\cite{Bertone:2005xz}. It was at the time believed
that IMBHs in the range $10^2-10^6$~M$\odot$ could have formed at the
place of DM overdensities and have increased the DM concentration (and
hence the annihilation rate) due to the gravitational
pull not strongly affected by the possible diluting effect of stellar activity. Such objects, yet undiscovered, could have been bright in
gamma-rays and dark in other wavelengths. Possible candidates were the
(back then) unidentified EGRET sources (EGRET is a precursor of
Fermi-LAT). The source 3EG1835 was, at that time, the brightest of the unidentified EGRET
sources. Unfortunately, the MAGIC data taken on that target were
affected by telescope technical problem that prevented a journal
publication. The results were discussed in conference~\cite{Doro:2007kta}.

Moving to non-galactic targets, galaxy clusters are
expected to host enormous amount of DM. Considering that 80\% of the
total mass content of the Universe is in the form of DM, and
considering the total mass of a galaxy clusters, for example the
Perseus cluster may host something like $10^{14-15}$~M$\odot$ in
DM. Answering the question whether galaxy clusters are optimal target
for DM searches is not trivial, because there are many processes at
work. On one side, that huge DM content may hint to large
concentrations at the barycenter, however, the same central region
maybe affected by strong outward winds of baryonic matter due to
standard astrophysical activity (GRBs, supernova, etc) which may
counteract the gravitational pressure and reduce the central DM
content. On the other side, in the case of annihilating DM, the contribution of
the DM substructures could be extremely high, with authors claiming a
factor of 10 to 1000 higher flux with respect to the ``smooth-halo''
case~\cite{SanchezConde:2011ap,Pinzke:2011ek,Gao:2011rf}. All in all, robust predictions in this case are very
hard to achieve. Less problematic is the case of the decaying DM in
galaxy clusters, because of the linear dependence on the DM
density (see Eq.~\ref{eq:dm_flux}). MAGIC preliminary results in single-telescope were presented
in \cite{Aleksic:2009ir}. In terms of annihilating DM, the constraints
were weak. In Sec.~\ref{sec:recent}, we will update this consideration
with the very-large Perseus campaigns performed with the stereoscopic MAGIC.

MAGIC is also an instrument capable of measuring cosmic ray
particles~\cite{Doro:2016scineghe}. Cosmic electrons, constituting
a few percents of the total cosmic ray flux arriving at the Earth,
initiate EAS totally similar to gamma-ray induced ones. The trick to
separate them is to consider only sky regions were no gamma-rays are
detected, and consider as control background that obtained with MC
simulations. Clearly the analysis is complex because of: $a)$ of the very
precise control of the MC-instrument matching required, and $b)$ 
of the fact that e-induced showers have an isotropic origin, and
images are more complex to reconstruct. The interest stems from the
fact that an excess in the positron spectrum was detected and
confirmed in the past years by several
instruments~\cite[and references therein]{Aguilar:2013qda}. This discrepancy can be explained by
standard astrophysical mechanisms like the emission from local
pulsar(s), or by secondary electrons produced in DM decay or
annihilation products. MAGIC can contribute significantly in an energy
range hardly at reach of satellite instruments such as AMS-II or
Fermi-LAT. MAGIC reported some preliminary results at
conferences~\cite{BorlaTridon:2011dk}. The follow-up study was
performed in stereoscopic mode, however, our data are strongly
dominated by systematic uncertainties, which make the all-electron spectrum only moderately
informative~\cite{Mallot:tevpa}.

\section{\label{sec:recent}Recent results}

Figure~\ref{fig:evolution} show the collection of some DM related
MAGIC results and their evolution. With the advent of the second telescope, the strategy to hunt DM with
MAGIC improved because:
\begin{enumerate}
\item As mentioned in Sec.~\ref{sec:early}, besides the obvious
  improvement from mono to stereo, the performance of the instrument received a
  strong boost along the years by instrumental upgrades;
\item It was clear that substantial effort in observation time
  should be devoted in order to gain interesting results. From
  Table~\ref{tab:targets} one can see that hundreds of hours were
  devoted to some targets;
\item The data reconstruction and analysis was optimized in several
  ways: $a)$ using a full likelihood approach that takes into account the DM
  spectral features, the instrument response function, the
  uncertainties in the background and the J-factor model, all resulting
  in a performance boost of a factor of 2 as well as preciser results~\cite{Aleksic:2012cp}, $b)$
  enlarging the search region to a larger range of DM masses, $c)$
  providing model-independent results for pure
  annihilation/decay channels instead of benchmarks models (that were
  based on LHC searches and lost
  importance as long as LHC was ruling them out) 
\end{enumerate}

As a result, the MAGIC exclusion curves have grown substantially
better. MAGIC devoted 160~h to the best DSG candidate (at that time):
Segue~1.
The Segue~1
stereo paper~\cite{Aleksic:2013xea} provided the absolute strongest upper
limits from DSG for DM particles above few hundreds GeV (at lower DM
masses, the results from Fermi-LAT are the most constraining). These
MAGIC results made into the PDG's Review of Particle Physics~\cite{DPG}. In addition,
MAGIC data were used for the first time in combination with Fermi-LAT data
to provide stronger constraints~\cite{Ahnen:2016qkx}. It is important
to stress that any further DSG observed with either Fermi-LAT or MAGIC
can be simply combined with previous observations thus resulting in a
global evolution toward stronger limits.

With the successful multi-year campaign on the Perseus Galaxy Cluster,
originally motivated also by cosmic ray astrophysics, MAGIC collected a
total of more than 250~h. These allowed to provide very strong constrain
on the Perseus core dynamics, and thermal-to-nonthermal radiation
balance~\cite{Ahnen:2016qkt}. The DM expectations in Perseus were
computed by several authors~\cite{SanchezConde:2011ap,Pinzke:2011ek,Gao:2011rf} and disagree of a factor of 100
one another due to different computation of the additional boost due
to substructures of the DM halo, as discussed in Sec.~\ref{sec:early}. For MAGIC, computing limits from
Perseus is hindered by the presence of a bright source at the center,
the radio galaxy NGC1275, and therefore an optimized analysis was
performed. The annihilating DM case can be constrained less
effectively than with DSG, however, for the decaying DM case, Perseus
is expected to deliver the strongest lower limits in decay lifetime for
DM particles above few hundreds GeV~\cite{Palacio:2015nza}. The full
paper is in preparation.



The Galactic Center was observed by MAGIC-mono for 25~h in
2006~\cite{Albert:2005kh}, however, the paper focused only on the
astrophysical interpretation. A larger campaign was made with
MAGIC-stereo in year 2012-16~\cite{Ahnen:2016crz}. The GC is
observable only at large zenith angles from MAGIC. This has a double
effect of largely increasing the energy threshold (because of stronger
extinction of the Cherenkov light in the atmosphere), but at the same
time increasing the effective area at high energies (due to a larger
footprint of Cherenkov photons at the ground). 
More than 70~h were
collected. The GC presents difficulties connected to the presence of
one or more bright and extended astrophysical targets at its
center, as well as diffuse gamma-ray emission. For this reason, we
expect MAGIC results to be competitive. The DM related paper is under
preparation.

Furthermore, about 50~h were devoted to the search of unidentified HE
targets, as described in Sec.~\ref{sec:early}. This time, the Fermi-LAT
all-sky catalogs were searched for stable, unassociated sources off
the galactic plane, optimal candidates to be DM overdensities. Results
were shown in conferences~\cite{Satalecka:2015yxa}. No detection was found. The paper is in
preparation too.

\section{\label{sec:future}Discussion and Outlook}

\begin{figure}[h!t]
  \centering
  \includegraphics[width=0.99\linewidth]{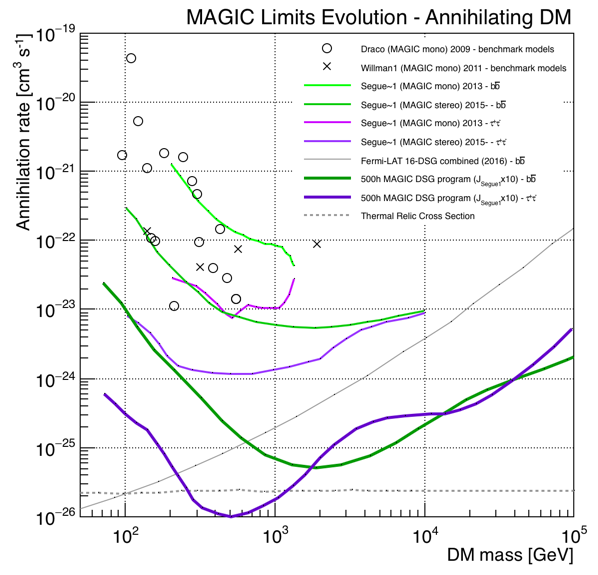}
  \caption{\label{fig:evolution}Collection of upper limits obtained
    with MAGIC. With Draco and Willman1 we used benchmark models. For
    Segue~1 we report results in the case of annihilating DM for the
    $b\bar{b}$ and $\tau^+\tau^-$ channels in case of MAGIC-mono,
    MAGIC-stereo and an extrapolation of MAGIC performance for a
    target 10 times brighter than Segue~1 observed for
    500~h. Additional Fermi-LAT~\cite{Ackermann:2015zua} exclusion curve and the thermal relic
    cross-section~\cite{Steigman:2012nb} are shown. Further details
    and discussion are given in the text.}
\end{figure}

The previous two sections have shown that MAGIC devoted a
substantial effort toward DM searches, which has increase from first
``snapshot'' observation to long-term observation campaigns. The capability of
extracting robust and optimized results has also increased, specially
by using a tailored full likelihood method as well as combining
results with other instruments. However, Fig.~\ref{fig:evolution}
shows that MAGIC upper limits are still some two orders of magnitude
above the thermal relic annihilation rate.

One could wonder whether we are too far from detection. The opinion of
the author is that the plot does not bear this information, due to the
following reasons: $a)$ there are several mechanisms for which a DM
particle satisfying all known constraints (e.g. the Sommerfeld effect,
which is expected because DM is cold) can have \sv larger than the
thermal value, $b)$ the presence of more than
one DM particle would translate into a higher relic annihilation rate
that what computed considering only 1 thermal relic,
$c)$ MAGIC and the other IACTA are the most sensitive instruments at a
TeV DM mass range, with no other competitors in the field, and as
such, regardless of the true nature of DM (to be discovered!), these
results constitute long-lasting unique information, $d)$
even null results from present and future direct detection or accelerator
experiments for DM, there is always a portion of the parameter space
only accessible to TeV instruments~\cite{Cahill-Rowley:2014cba} $e)$
Fig.~\ref{fig:evolution} also plots an estimation of what exclusion
curve MAGIC could provide if observing a DSG for 500~h (this can be
done over several years) with an astrophysical factor 10 times larger
than of Segue~1 (as in Ref.~\cite{Ahnen:2016qkx})\footnote{The reader
  should be warned that this is a simple rescaling of the
  Ref.~\cite{Ahnen:2016qkx} curve, which refers to a specific dataset
  (with its fluctuations) and does not make a full computation using
  the expected sensitivity, for simplicity. The order of magnitude of
  the result is still valid.}. These curves would
be then very close to the thermal annihilation rate curve, even
crossing it in the hard $\tau^+\tau^-$ annihilation
channel. Considering that such a target may exist, this is definitely
a motivating point for MAGIC to maintain this effort.

Consequently, MAGIC will continue pursuing DM searches. The most promising targets
remain the DSGs, because of their clean environment, and strong
expected signal, which - if detected - would constitute a clear
detection compared to targets with strong astrophysical
background. This is also justified by the fact that many new DSGs have been
discovered in the recent times~\cite{Bechtol:2015cbp,Koposov:2015cua,Laevens:2015una,Laevens:2015kla,Kim:2015xoa,Martin:2015xla,Kim:2015ila,Drlica-Wagner:2015ufc}, and more are
expected in future campaigns. Some of these are also modeled with
extremely large DM content (expected from simulations, see Fig.3
of~\cite{Hutten:2016jko}).  All future MAGIC DSGs observation could be
combined together as done in~\cite{Ahnen:2016qkx}.

\section{Conclusions}
\vspace{-3mm}
In this contribution, we have shown the evolution of MAGIC results for
dark matter searches at different target of interests. The dedication
of hundreds of hours allowed MAGIC to place the most robust upper
limits above few hundreds GeV on annihilating dark matter particle
models through the observation of highly dark matter dominated
satellite galaxies. The long campaign on Perseus will allow to put
similar strong constraints on the lifetime of decaying dark
matter. MAGIC continues its dark matter program in order to provide
legacy results before the times of CTA. 

\begin{small}
\end{small}

\bigskip 

\end{document}